\documentclass[cameraready]{Interspeech}

\title{ENHANCING AUDIO CAPTIONING WITH AUXILIARY AUDIOSET SEMANTICS}

\author[affiliation={1}, equalcontribution]{Shubham}{Gupta}
\author[affiliation={1}, equalcontribution]{Adarsh}{Arigala}
\author[affiliation={1}]{Sri Rama Murty}{Kodukula}
 
\address{
$^1$ Speech Information and Processing Lab,\\
Indian Institute of Technology Hyderabad, India
}

\email{guptashubham0318@gmail.com, arigalaadarsh780@gmail.com,  ksrm@ee.iith.ac.in}


\usepackage{comment}
\usepackage{multirow}
\usepackage{makecell}
\usepackage{subcaption}
\usepackage{amssymb}
\usepackage{pifont}

\begin{document}

\maketitle

\begingroup
\renewcommand\thefootnote{}
\footnotetext{Source code and pretrained models are available at:
\url{https://github.com/ArigalaAdarsh/Enhancing-AAC-AudioSet-Semantics.git}}
\endgroup
\begin{abstract}
Automatic Audio Captioning (AAC) seeks to generate natural language descriptions of complex acoustic scenes, bridging auditory perception and language understanding. However, word-selection indeterminacy and increasing reliance on large-scale sequence-to-sequence or LLM-based models limit practical deployment. We propose a resource-efficient AAC framework that explicitly grounds caption generation in auxiliary AudioSet semantics. Frame-level acoustic representations extracted using a ConvNeXt encoder are augmented with top-$K$ predicted AudioSet keywords, providing structured contextual cues for decoding. A compact six-layer BART-style decoder conditions on this joint acoustic-semantic representation, enabling caption generation without LLM-scale decoding. The proposed design balances semantic grounding and computational efficiency within a compact architecture. Evaluations on Clotho V2 and AudioCaps confirm competitive caption quality under practical deployment constraints. 

\end{abstract}

\section{Introduction}
\label{sec:intro}
Automated Audio Captioning (AAC) transforms raw audio signals into structured textual descriptions, enabling applications in multimedia retrieval, security surveillance, and assistive technologies. Unlike traditional audio tagging, AAC must model not only discrete acoustic events but also broader scene-level context, including environmental ambience, object interactions, and perceptual attributes such as “muffled speech” or “distant traffic.” This cross-modal reasoning requires jointly capturing fine-grained acoustic details and higher-level semantic structure to produce meaningful and interpretable captions.

Despite recent progress, AAC remains challenging due to limited paired audio-text datasets, overlapping sound events, and inconsistent captions that introduce noise into supervision. Capturing scene-level context further increases complexity, while existing models often rely on shallow decoders or pretrained language models that may struggle with domain-specific acoustic nuances. Moreover, AAC systems face significant indeterminacy in word selection: a single acoustic event or scene can be described using various words or phrases, leading to a combinatorial explosion of possible captions and complicating the learning process.

Recent advances leverage large sequence-to-sequence architectures and LLM-based decoding strategies, including prefix tuning and post-hoc caption refinement, to improve fluency and semantic richness. While effective, these approaches typically require substantial model capacity and computational resources, limiting their suitability for resource-constrained or real-time deployment. Other works incorporate auxiliary tags or topic priors to improve grounding, but often rely on heavy decoders or complex multi-encoder pipelines, and rarely analyze the tradeoff between semantic guidance and model efficiency in a controlled manner.

To address these gaps, we propose a resource-efficient AAC framework that explicitly grounds caption generation in predicted AudioSet semantics. A ConvNeXt-based audio encoder extracts frame-level acoustic features, which are augmented with top-$K$ predicted AudioSet keywords. These keywords are embedded and fused with acoustic representations and supplied to a compact six-layer BART-style decoder that generates captions conditioned on this joint acoustic-semantic context. By directing the decoder with explicit semantic cues, our design reduces the decoder’s capacity requirements while improving semantic coherence and cross-domain robustness. The main contributions of this work are:
\begin{enumerate}
    \item We introduce a simple, effective fusion of frame-level acoustic features with top-$K$ AudioSet keyword embeddings that explicitly grounds caption generation and reduces word-selection indeterminacy.
    \item We propose a compact six-layer BART-style decoder and provide ablations demonstrating that, when guided by semantic keywords, reduced decoder capacity attains a favorable efficiency-quality Pareto frontier.
    \item We systematically analyze the role of semantic tags through tag-only, audio-only, and fusion ablations, highlighting complementary gains from joint modeling.
    \item We evaluate comprehensively on Clotho V2 and AudioCaps, including SPIDEr and FENSE metrics, cross-dataset analysis, and sensitivity studies on the number of keywords ($K$).
\end{enumerate}


\begin{figure*}[!t]  
    \centering
    \includegraphics[width=\linewidth]{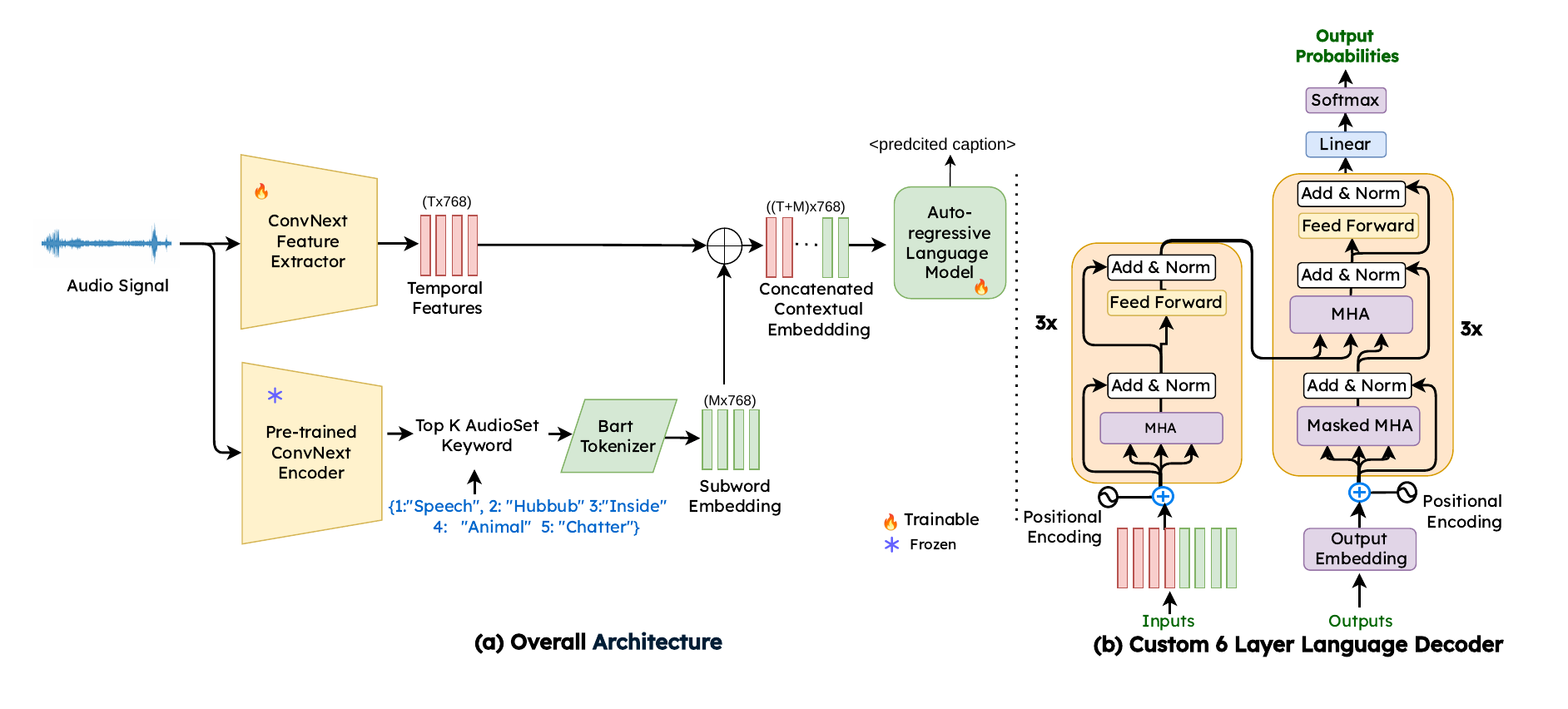}
    \captionsetup{font=small}
    \caption{Overview of the proposed AAC architecture.}
    \label{fig:architecture} 
\end{figure*}

\section{Related Works and Baselines}  
\label{sec:related_work}
Prior studies have explored auxiliary semantic cues, encoder fusion strategies, and advanced language modeling to address data scarcity and improve semantic grounding in AAC. Koizumi et al. \cite{3} guided caption generation by leveraging the most frequent words from Clotho captions, which improved semantic relevance but lacked fine-grained acoustic grounding and contextual flexibility. Eren et al. \cite{4} proposed a BiGRU encoder-decoder architecture that fused PANNs audio features with subject-verb embeddings, demonstrating the benefit of combining structural semantics with acoustic information. Gontier et al. \cite{1} conditioned a BART-based decoder on PANNs \cite{kong2020panns} features and YAMNet \cite{Hershey2017YAMNet} tags via topic modeling, achieving strong performance but at the cost of increased architectural complexity and computational demand.

More recent approaches have shifted toward leveraging large language models (LLMs) and large audio-language models (LALMs). Kim et al. \cite{2} introduced GPT-based decoding via prefix tuning, enabling adaptation to AAC with reduced fine-tuning cost but still relying on large-scale pretrained language models. Pengi \cite{Pengi} extended this paradigm by training a large audio-language model on diverse audio-text tasks, demonstrating strong zero-shot and cross-task transfer performance across benchmarks. Recent DCASE challenge submissions employ caption refinement strategies, multi-encoder fusion, LLM-based re-ranking, or post-hoc summarization \cite{2,9}, improving fluency and semantic richness while increasing computational complexity. While these methods advance caption quality, many rely on heavy decoders, large pretrained language models, or complex multi-encoder pipelines. In contrast, we focus on grounding caption generation in predicted AudioSet semantics while systematically analyzing the tradeoff between semantic guidance and decoder capacity within a compact, deployment-oriented architecture.

\begin{table*}[!t]
\centering
\caption{Evaluation on Clotho. Methods are grouped based on training source. 
(i) Clotho, 
(ii) AudioCaps and 
(iii) Large pretrained or prompt-based methods.
$\dagger$ Results are directly quoted from the original papers, or we trained the models using the publicly available codes otherwise. Best results are shown in \textbf{bold}, and the second-best results are \underline{underlined} within settings (i) and (ii) only.}
\renewcommand{\arraystretch}{1} 
\setlength{\tabcolsep}{4pt}
\begin{tabular}{lccccccccc}
\hline
Training Set & Method &  
BLEU\textsubscript{1} & BLEU\textsubscript{4} & METEOR & 
ROUGE\textsubscript{L} & CIDEr & SPICE & 
SPIDEr & FENSE \\ \hline

\multirow{6}{*}{(i) Clotho} 
& $\dagger$Gontier et al.~\cite{1}  & 0.506 & 0.134 & 0.148 & 0.338 & 0.278 & 0.092 & 0.185 & - \\
& Kim et al.~\cite{2}               & 0.560 & \underline{0.160} & 0.170 & 0.378 & \underline{0.392} & \underline{0.118} & \underline{0.255} & 0.336 \\
& DCASE 2023                        & 0.565 & 0.151 & 0.171 & 0.371 & 0.388 & 0.117 & 0.252 & \underline{0.437} \\
& $\dagger$Koizumi et al.~\cite{3}  & 0.521 & 0.107 & 0.149 & 0.342 & 0.258 & 0.097 & 0.177 & - \\
& $\dagger$Eren et al.~\cite{4}     & \underline{0.590} & 0.140 & \textbf{0.220} & \textbf{0.450} & 0.280 & - & - & - \\
& Ours                     & \textbf{0.605} & \textbf{0.183} & \underline{0.182} & \underline{0.393} & \textbf{0.447} & \textbf{0.123} & \textbf{0.285} & \textbf{0.478} \\ \hline

\multirow{4}{*}{(ii) AudioCaps}
& DCASE 2023                        & 0.296 & 0.060 & 0.102 & 0.250 & 0.157 & 0.064 & 0.110 & - \\
& $\dagger$Gontier et al.~\cite{1}  & 0.309 & 0.034 & 0.098 & 0.233 & 0.112 & 0.046 & 0.079 & - \\
& Kim et al.~\cite{2}               & \underline{0.342} & \underline{0.065} & \underline{0.112} & \underline{0.276} & \underline{0.192} & \underline{0.074} & \underline{0.133} & \underline{0.325} \\
& Ours                     & \textbf{0.374} & \textbf{0.078} & \textbf{0.122} & \textbf{0.282} & \textbf{0.225} & \textbf{0.076} & \textbf{0.142} & \textbf{0.419} \\ \hline

\multirow{2}{*}{\shortstack{(iii) Large Pretrained \\ \& Prompt-based}}
& Pengi~\cite{Pengi}                & 0.556 & 0.144 & 0.166 & 0.375 & 0.400 & 0.126 & 0.260 & 0.488 \\
& Keyword→LLaMA-2-7B  & 0.317 & 0.020 & 0.117 & 0.227 & 0.118 & 0.073 & 0.095 & 0.445 \\ \hline

\end{tabular}
\label{tab:results_clotho}
 
\end{table*}

\begin{table*}[!t]
\centering
\caption{Evaluation on AudioCaps. Methods are grouped based on training source. 
(i) Clotho, 
(ii) AudioCaps and 
(iii) Large pretrained or prompt-based methods.
$\dagger$ Results are directly quoted from the original papers, or we trained the models using the publicly available codes otherwise. Best results are shown in \textbf{bold}, and the second-best results are \underline{underlined} within settings (i) and (ii) only.}
\renewcommand{\arraystretch}{1} 
\setlength{\tabcolsep}{4pt}
\begin{tabular}{lccccccccc}
\hline
Training Set & Method & BLEU\textsubscript{1} & BLEU\textsubscript{4} & METEOR & ROUGE\textsubscript{L} & CIDEr & SPICE & 
SPIDEr & FENSE \\ \hline

\multirow{5}{*}{(i) AudioCaps}
& DCASE 2023 & 0.581 & 0.189 & 0.185 & 0.427 & 0.552 & 0.139 & 0.345 & - \\
& $\dagger$Gontier et al.~\cite{1} & 0.699 & 0.266 & 0.241 & 0.493 & \underline{0.753} & 0.176 & \underline{0.465} & - \\
& $\dagger$Eren et al.~\cite{4} & 0.710 & 0.230 & \textbf{0.290} & \textbf{0.590} & 0.750 & - & - & - \\
& Kim et al.~\cite{2} & \underline{0.713} & \textbf{0.309} & 0.240 & 0.503 & 0.733 & \underline{0.177} & 0.455 & \underline{0.357} \\
& Ours & \textbf{0.716} & \underline{0.302} & \underline{0.246} & \underline{0.507} & \textbf{0.780} & \textbf{0.180} & \textbf{0.470} & \textbf{0.615} \\ \hline

\multirow{4}{*}{(ii) Clotho}
& DCASE 2023 & 0.421 & 0.064 & 0.135 & 0.314 & 0.187 & 0.079 & 0.133 & - \\
& $\dagger$Gontier et al.~\cite{1} & 0.425 & 0.061 & 0.128 & 0.298 & 0.147 & 0.060 & 0.104 & - \\
& Kim et al.~\cite{2} & \underline{0.449} & \underline{0.084} & \underline{0.144} & \underline{0.330} & \underline{0.211} & \underline{0.083} & \underline{0.147} & \underline{0.283} \\
& Ours & \textbf{0.519} & \textbf{0.120} & \textbf{0.174} & \textbf{0.366} & \textbf{0.298} & \textbf{0.118} & \textbf{0.208} & \textbf{0.437} \\ \hline

\multirow{2}{*}{\shortstack{(iii) Large Pretrained \\ \& Prompt-based}}
& Pengi~\cite{Pengi} & 0.515 & 0.106 & 0.170 & 0.368 & 0.398 & 0.113 & 0.256 & 0.494 \\
& Keyword→LLaMA-2-7B & 0.327 & 0.023 & 0.148 & 0.263 & 0.118 & 0.088 & 0.103 & 0.535 \\ \hline

\end{tabular}
\label{tab:results_audiocaps}
 
\end{table*}

The proposed framework consists of three key components: (i) an Audio Encoder that extracts temporal acoustic representations, (ii) a Keyword Module that provides semantic context through predicted AudioSet labels, and (iii) a Language Decoder that generates descriptive captions by integrating acoustic and semantic features as shown in Fig. \ref{fig:architecture}.

\subsection{Audio Encoder}
We employ a ConvNeXt-Tiny encoder \cite{6} pretrained on AudioSet \cite{7} to extract acoustic representations. Given an input audio signal $x$, the encoder produces frame-level embeddings  $\mathbf{H}_a \in \mathbb{R}^{T \times d}$, where $T$ denotes the number of temporal frames and $d$ is the embedding dimension.

\subsection{Keyword Module}
To provide semantic grounding, we employ a separate ConvNeXt-Tiny classifier pretrained on AudioSet and kept frozen. Given an input audio signal $x$, it produces logits over the 527 AudioSet classes, from which the top-$K$ classes $\mathcal{C}_K$ are selected. Each selected class label is tokenized using the BART-base tokenizer. Let $M$ denote the total number of resulting subword tokens across all $K$ labels. The corresponding embeddings are represented as 
$\mathbf{K} = [\mathbf{k}_1, \mathbf{k}_2, \ldots, \mathbf{k}_M] \in \mathbb{R}^{M \times d}$. These keyword embeddings are concatenated with the
acoustic representations along the temporal dimension: $\mathbf{H}_f = [\mathbf{H}_a ; \mathbf{K}] \in \mathbb{R}^{(T + M) \times d}$. This fused representation provides explicit semantic context for caption generation.

\subsection{Language Decoder}
The Language Decoder is a compact, custom six-layer BART-style architecture comprising multi-head attention mechanisms. Specifically, the decoder consists of three encoder and three decoder Transformer layers. To tokenize target captions, we employ a pre-trained BART-base tokenizer, leveraging its rich contextual understanding from large-scale text corpora. At each decoding step $t$, the decoder generates the next token $y_t$ by modeling the conditional probability $p(y_t \mid y_{<t}, \mathbf{H}_f)$ as \[y_t = \arg\max_{w \in \mathcal{V}} \; p(w \mid y_{<t}, \mathbf{H}_f)\]
where $\mathcal{V}$ is the vocabulary space. The decoder is trained in an auto-regressive manner to minimize the cross-entropy loss between predicted tokens and ground-truth captions, i.e., $\mathcal{L}_{\mathrm{CE}} = -\sum_{t=1}^{L} \log p(y_t \mid y_{<t}, \mathbf{H}_f)$, where $y_t$ is the ground-truth token at time step $t$ and $L$ is the target caption length. 

During inference, beam search with a beam width of 5 is employed to generate fluent and contextually coherent captions. By reducing BART to a compact six-layer design, the decoder lowers model complexity, memory footprint, and training time, while effectively leveraging fused acoustic and semantic embeddings for high-quality audio captioning.

\section{Experiments and Results}

\subsection{Datasets}
We conducted experiments on two audio captioning benchmarks: AudioCaps \cite{12} and Clotho-V2 \cite{8}. AudioCaps contains 50,000 ten-second clips sourced from AudioSet, each paired with one caption for training and five captions per clip for evaluation. Clotho-V2 comprises 4,981 audio samples ranging from 15 to 30 seconds, each annotated with five captions that are typically 8 to 20 words long, offering rich linguistic diversity. For both datasets, we follow the standard protocols for training, validation, and test splits as prescribed in the original papers, ensuring fair and consistent evaluation across methods.             

\subsection{Implementation Details}
All audio from the Clotho and AudioCaps datasets was resampled to 32\,kHz and transformed into log-Mel spectrograms (224 Mel bins) computed with a 32\,ms window and 10\,ms hop. Training was performed for 20 epochs with a batch size of 4 and gradient accumulation of 2, using the AdamW optimizer with a learning rate of $1\times10^{-5}$ and $K=5$ predicted keywords from AudioSet. The ConvNeXt audio encoder was fine-tuned to enhance feature extraction for captioning task. The language decoder consists of three Transformer encoder and three decoder layers and is trained from scratch. Each layer uses a model dimension of 768, eight attention heads, a feed-forward hidden size of 2048 with GELU activation, and a dropout rate of 0.1.

\begin{table*}[!t] 
\centering
\caption{Impact of keyword guidance and decoder architecture on captioning 
performance on the Clotho dataset. \ding{51} indicates the use of AudioSet 
keyword conditioning and \ding{55} denotes its absence.}
\renewcommand{\arraystretch}{0.9} 
\begin{tabular}{lcccccccc}
\hline
\makecell{Language \\ Decoder} & \makecell{Keyword\\ Guidance} & Params (M) & FLOPs (G) & BLEU\textsubscript{1} & METEOR & ROUGE\textsubscript{L} & CIDEr & SPIDEr \\
\hline
BART-Base & \ding{55} & 169 & 95.11 & 0.555 & 0.170 & 0.368 & 0.382 & 0.248  \\
BART-Large & \ding{55} & 437 & 124.18 & 0.478 & 0.155 & 0.329  & 0.258 & 0.179 \\
Ours & \ding{55}  & 110 & 88.55 & 0.588 & 0.175 & 0.387 & 0.402 & 0.260  \\ \hline
BART-Base & \ding{51}  & 169 & 95.11 &  0.563 & 0.172 & 0.371  & 0.414 & 0.267 \\
BART-Large & \ding{51} & 437 & 124.18 & 0.517 & 0.174 & 0.357 &  0.363 & 0.241 \\
Ours & \ding{51} & 110 & 88.55 & \textbf{0.605} &  \textbf{0.182} & \textbf{0.393} &\textbf{0.447} & \textbf{0.285}  \\
\hline
\end{tabular}
\label{tab:bart_ablation}
 
\end{table*}

\subsection{Results and Comparison}  
We compare our proposed model with competing baselines under two evaluation paradigms: (i) in-domain evaluation, where training and test sets originate from the same dataset (e.g., both from Clotho), and (ii) cross-domain evaluation, where the model is trained on one dataset and evaluated on another (e.g., training on Clotho and testing on AudioCaps). All methods are evaluated using metrics widely adopted in captioning tasks, including BLEU \cite{26}, METEOR \cite{27}, ROUGE\textsubscript{L}\cite{28}, CIDEr \cite{29}, SPICE \cite{30}, SPIDEr \cite{31}, and FENSE \cite{fense}.

As summarized in Table \ref{tab:results_clotho} and Table \ref{tab:results_audiocaps}, under the in-domain setting, our model demonstrates competitive performance across most metrics compared with prior supervised approaches. On Clotho, the proposed model achieves a SPIDEr score of 0.285 and FENSE of 0.478, outperforming all competing compact methods. Improvements are also consistent across BLEU and CIDEr, indicating that explicit AudioSet semantic grounding reduces caption ambiguity and improves lexical precision. A similar trend is observed on AudioCaps, where the model attains SPIDEr 0.470 and FENSE 0.615.

Under cross-domain setting, in both scenarios, our approach consistently outperforms competing supervised baselines across nearly all metrics. Competing methods trained on a single distribution exhibit substantial metric degradation when evaluated on the other dataset, indicating susceptibility to dataset-specific patterns. In contrast, our model maintains more stable performance across domains. Notably, AudioCaps (sourced from YouTube) and Clotho (sourced from Freesound) differ significantly in noise levels, compression artifacts, and event density; however, cross-dataset evaluation shows smaller performance degradation compared to baselines, indicating reduced sensitivity to dataset bias.

For additional context, we also report results from large pretrained audio-language or prompt-based approaches, including Pengi \cite{Pengi} and a keywords→LLaMA-2-7B \cite{llama2} pipeline. While Pengi benefits from large-scale pretraining and multi-task learning and reports SPIDEr 0.260 on Clotho and 0.256 on AudioCaps, our model achieves SPIDEr 0.285 and 0.470 despite using a substantially smaller architecture and fewer parameters, demonstrating a favorable efficiency–quality trade-off enabled by explicit semantic grounding.

The Keyword → LLaMA-2-7B baseline, in which top-K predicted AudioSet labels are fed directly to LLaMA-2 LLM and prompted to generate plausible captions without any audio feature conditioning, achieves SPIDEr 0.095 on Clotho and 0.103 on AudioCaps. The pronounced gap between this pipeline and the proposed model confirms that keyword guidance alone is insufficient for accurate caption generation when acoustic context is not jointly modeled.


\subsection{Ablation studies}
\label{sec:Ablation}
 
\noindent \textbf{Effect of Keyword Set Size ($K$) on Captioning Performance}\\ 
We study the effect of varying the number of predicted AudioSet keywords ($K$) used for semantic grounding. As shown in Table~\ref{tab:topk_audiowords}, $K=5$ yields the best overall performance, achieving the highest BLEU-1 (0.605), METEOR (0.182), CIDEr (0.447), and SPIDEr (0.285). Increasing $K$ to 10 leads to consistent degradation in CIDEr and SPIDEr, while further expansion to $K=15$ does not provide additional gains. This trend highlights a tradeoff between semantic coverage and noise injection: smaller $K$ values emphasize high-confidence predictions that effectively constrain the caption space, whereas larger $K$ introduces lower-confidence classes that dilute the semantic signal and increase ambiguity during decoding. Although increasing $K$ may marginally broaden lexical diversity, it reduces consensus-based metrics, indicating diminished caption precision. Based on these observations, we adopt $K=5$ for all subsequent experiments.

\begin{table}[!h]
\centering
\caption{Effect of varying the number of predicted keywords ($K$) on captioning performance on the Clotho dataset using the proposed method. Bold indicates the best result for each metric.}
\renewcommand{\arraystretch}{1 } 
\setlength{\tabcolsep}{5pt}
\begin{tabular}{cccccccc}
\hline
\textbf{\makecell{$K$}} & BLEU\textsubscript{1} & METEOR & ROUGE\textsubscript{L} & CIDEr & SPIDEr \\
\hline
1 & 0.603 & 0.181 &   0.395 & 0.436 & 0.281    \\
2 & 0.599 &  0.181 & 0.395 & 0.429  & 0.276 \\
3 & 0.604 & 0.180 &  \textbf{0.396} & 0.434 & 0.279 \\ 
4 & 0.600 & 0.179 & 0.393 & 0.429 & 0.277 \\
5  &  \textbf{0.605}   & \textbf{0.182} & 0.393 & \textbf{0.447} &  \textbf{0.285} \\
10 & 0.599  & 0.179 & 0.393 & 0.433 &   0.279 \\
15 & 0.596   & 0.180 & 0.388 & 0.427 &   0.275 \\
\hline
\end{tabular}
\label{tab:topk_audiowords}
 
\end{table}

\noindent\textbf{Impact of Keyword Guidance} To analyze the impact of semantic guidance and decoder architecture, we compare three language decoder variants: BART-Base, BART-Large, and our compact six-layer decoder under two settings: with and without AudioSet keyword conditioning (Table~\ref{tab:bart_ablation}). Incorporating predicted keywords consistently improves caption quality across all decoders, highlighting the value of explicit semantic cues. For example, our decoder improves from BLEU\textsubscript{1} 0.588 to 0.605, CIDEr 0.402 to 0.447, and SPIDEr 0.260 to 0.285 when keyword guidance is enabled. Similar improvements are observed for BART-based decoders, indicating that AudioSet tags provide complementary high-level context that reduces caption ambiguity and better aligns captions with salient audio events. Despite being significantly smaller, our six-layer decoder outperforms both BART-Base and BART-Large across most metrics. This suggests that once semantic cues are available, increasing decoder capacity yields diminishing returns and may introduce excessive language priors when training on relatively small audio captioning datasets.

\begin{figure}[!h]
    \centering
    \includegraphics[width=\linewidth]{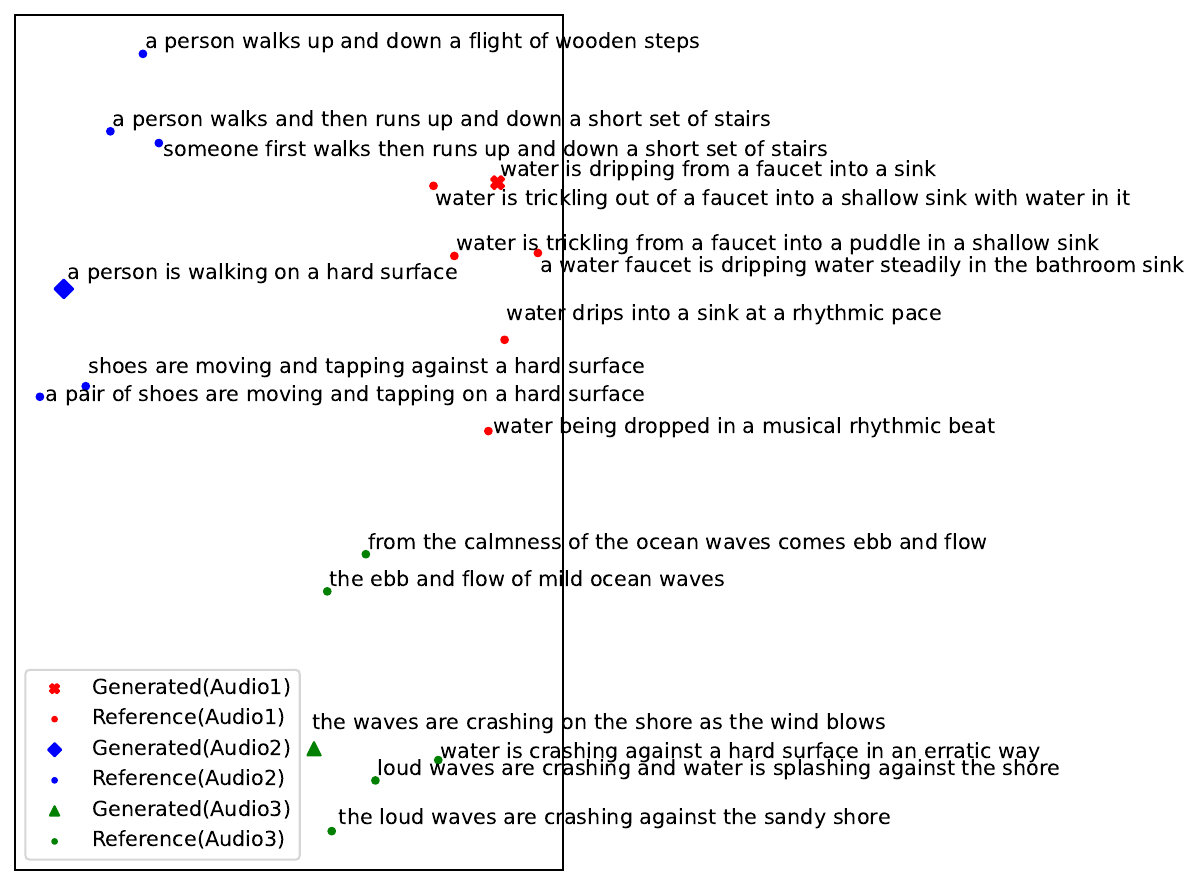}
    \caption{ t-SNE visualization of caption embeddings demonstrating semantic similarity between ground truth and model-predicted captions.}
    \label{fig:ours_tsne}
    \vspace{-0.5cm}
\end{figure}

\subsection{Qualitative results}
\label{sec:Qualitative results}
We visualize the semantic alignment between predicted and reference captions using t-SNE. All captions are embedded using DistilBERT \cite{distilbert} (distilbert-base-nli-stsb-mean-tokens, 768-dim). As shown in Fig.~\ref{fig:ours_tsne}, predicted captions cluster closely with their corresponding ground-truth captions for each audio sample, indicating strong semantic consistency and meaningful audio–text alignment.

\section{Conclusion}
\label{sec:conclusion}
This work presents a balanced framework for automated audio captioning that integrates AudioSet semantic cues with a ConvNeXt-based encoder and a lightweight six-layer BART-style decoder. By incorporating predicted AudioSet keywords, the proposed approach helps mitigate word-selection indeterminacy and improves semantic alignment between audio events and generated captions. Experiments on Clotho V2 and AudioCaps demonstrate competitive performance with reduced computational complexity and strong cross-domain generalization. Future work will explore scaling to larger datasets, multimodal extensions, and improved robustness in real-world acoustic environments.

\bibliographystyle{IEEEtran}
\bibliography{mybib}

\end{document}